\documentclass[12pt]{iopart}

\usepackage{epsfig}
\usepackage{graphicx}
\begin{document}

\title{Feasibility of 300 km Quantum Key Distribution with Entangled States}

\author{Thomas Scheidl$^1$, Rupert Ursin$^{1}$, Alessandro Fedrizzi$^1$, Sven Ramelow$^1$, Xiao-Song Ma$^1$, Thomas Herbst$^2$, Robert Prevedel$^2$, Lothar Ratschbacher$^1$, Johannes Kofler$^1$, Thomas Jennewein$^1$ and Anton Zeilinger$^{1,2}$}

\address{$^1$ Institute for Quantum Optics and Quantum Information, Austrian Academy of Sciences, Vienna\\
$^2$ Faculty of Physics, University of Vienna}

\ead{zeilinger-office@univie.ac.at}
\begin{abstract}
A significant limitation of practical quantum key distribution (QKD) setups is currently their limited operational range. It has recently been emphasized \cite{Ma07} that entanglement-based QKD systems can tolerate higher channel losses than systems based on weak coherent laser pulses (WCP), in particular when the source is located symmetrically between the two communicating parties, Alice and Bob. In the work presented here, we experimentally study this important advantage by implementing different entanglement-based QKD setups on a 144~km free-space link between the two Canary Islands of La Palma and Tenerife. We established three different configurations where the entangled photon source was placed at Alice's location, asymmetrically between Alice and Bob and symmetrically in the middle between Alice and Bob, respectively. The resulting quantum channel attenuations of 35~dB, 58~dB and 71~dB, respectively, significantly exceed the limit for WCP systems \cite{Ma07}. This confirms that QKD over distances of 300~km and even more is feasible with entangled state sources placed in the middle between Alice and Bob.
\end{abstract}

\maketitle

\section{Introduction}
Quantum cryptography, which promises to solve the problem of secure key distribution for the encryption of messages, is the most mature technical application in the field of quantum information and quantum communication. Current QKD architectures can be broadly categorized into systems based either on weak coherent laser pulses (WCP) \cite{Hwang03,peng07,hasegawa07,Schmitt07}, on continuous variables \cite{Ralph99,Hillery00,ralph00,lodewyck07,qi07} or on entanglement \cite{Ekert91,Bennett92,honjo08}. On the commercial market, WCP systems have been available for a while \cite{magicQ,quantique}. Most recently a QKD network was successfully demonstrated in Vienna (Peev \emph{et al.} in the same issue), including a fully automatic entanglement-based QKD system, operating reliably with installed telecom fibers (Treiber \emph{et al.} in the same issue \cite{treiber09}).
\par
An important benchmark for a QKD system is the secure key rate that can be achieved for a given quantum channel attenuation. Due to absorptive losses in the communication channels and detector imperfections, the distance/attenuation over which a secure key can still be generated is limited for all QKD systems. The experimental method which presently offers the best performance in high loss regimes are symmetric entanglement-based systems. This was recently shown by X-F. Ma \emph{et al.} \cite{Ma07}, where the authors conclude that state-of-the-art pulsed entanglement based QKD systems with the source placed symmetrically in the middle between the receivers can cover up to twice as much attenuation as WCP systems. In their particular example, the maximal attenuation was evaluated to be an astonishing 70~dB (in the case of optimized mean photon number in the limit of an infinite number of key bits) when using experimental parameters as in~\cite{ursin07}. Comparatively, a WCP system based on the same parameters can cover only up to 35~dB attenuation.
\par
The quantum channels in a future global quantum communication networks will mostly consist of optical fibers which are already widely installed. As an alternative, free-space connections  will allow to quickly build up connections between parties with direct line-of-sight \cite{Bennett92d,jacobs96,Buttler98a,Buttler00a,kurtsiefer02,kurtsiefer02b,ursin07,Erven08,peloso08}. Additionally, orbital free-space links, e.g. satellite-to-ground links or inter-satellite links, will allow the efficient global interconnection of regional quantum networks \cite{Buttler98c,ursin08b}. The attenuation expected for a single link ground connection from a satellite is at least 30~dB, and its feasibility has been shown in first ground-based tests \cite{Schmitt07,ursin07}. In the more demanding two-link satellite scenario, QKD systems will have to cope with 60~dB attenuation.
\par
In this work, we experimentally test the performance of entanglement-based QKD in an attenuation range from 35~dB to 70~dB over a distance of 144~km. We demonstrate the feasibility of entanglement-based QKD in loss regimes, where secure communication is no longer possible using WCP systems. With respect to the channel symmetry, we show that the obtained secure key rates confirm the predicted advantage of the symmetric scenario over commonly used asymmetric systems and we compare our results with the theoretical model by X.-F. Ma \emph{et al.} \cite{Ma07}.

\subsection{Entanglement-based QKD}
The most commonly used entanglement-based QKD scheme is the BBM92 protocol \cite{Bennett92}, where it was shown to be equivalent to the original BB84 scheme \cite{Bennett84} under ideal conditions: Ideally, a source generates polarization entangled photons in the state
\begin{equation}\label{psiminus}
    |\psi^-\rangle=\frac{1}{\sqrt{2}}\left(|H\rangle_a|V\rangle_b-|V\rangle_a|H\rangle_b\right),
\end{equation}
where $|H\rangle$ ($|V\rangle$) denotes horizontal (vertical) polarization. The photons in modes \emph{a} and \emph{b} are sent through quantum
channels to the two communicating parties, Alice and Bob. Both perform measurements on the incoming photons in one out of two randomly chosen, complementary bases. Let's assume that they chose between the $|H,V\rangle$-basis and the $|P,M\rangle$-basis, with $|P\rangle=\frac{1}{\sqrt{2}}(|H\rangle+|V\rangle)$ and $|M\rangle=\frac{1}{\sqrt{2}}(|H\rangle-|V\rangle)$. Alice and Bob individually record their measurement outcomes, including the information about the measurement basis. Assigning the binary value ``0'' to the results $|H\rangle$ and $|P\rangle$ and the value ``1'' to the results $|V\rangle$ and $|M\rangle$, each of the observers obtains a completely random bit string, the \emph{raw key}. The \emph{sifted key} is gained after basis reconciliation, where the raw key is reduced by the basis reconciliation factor of $2$. This is due to the fact that Alice and Bob discard those events where they have accidentally chosen different bases. Since the quantum state (\ref{psiminus}) received by Alice and Bob was entangled, the retained measurement outcomes are perfectly anti-correlated, i.e. Alice's and Bob's sifted keys are perfectly inverse to each other. This key is then further used to encode a message which can be transmitted over a public channel.
\par
The security against an eavesdropper, Eve, is guaranteed by the laws of quantum mechanics. Any attempt by Eve to gain information on the transmitted qubits will inevitably reduce the entanglement and introduce errors in the sifted key. The quantum bit error ratio (QBER) $q$ is constantly monitored by Alice and Bob by comparing small parts of their keys. As long as the QBER stays below a certain value, classical error correction and privacy amplification protocols can be used to distill an unconditional secure key.
\par
In real-world QKD experiments, errors will most probably be caused by imperfections in the setup rather than by Eve. However, for unconditional security, all errors need to be treated as if coming from an attempted eavesdropping attack. The next section is devoted to a theoretical analysis how the various imperfections affect the QBER and in consequence limit the achievable secure key rate.

\subsection{Theoretical error model}
In practical QKD setups, most errors will actually originate from experimental imperfections, e.g. non-perfect entanglement and higher order photon emissions at the source, noisy quantum channels, imperfect polarization analyzers and photon detectors. A direct estimate of the expected QBER in a quantum optics QKD experiment can be obtained by measuring the total quantum correlation visibility $V_{tot}$, which has a simple relation to $q$:
\begin{equation}\label{QBER}
q =\frac{1-V_{tot}}{2}
\end{equation}
and can be obtained from the maxima, $N_{max}$, and minima, $N_{min}$, of the observed coincidences:
\begin{equation}
V_{tot}=\frac{N_{max}-N_{min}}{N_{max}+N_{min}}.
\end{equation}
Given that all these parameters are experimentally accessible, one can model the performance of a QKD system. First, a finite coincidence time window $\tau_c$, limited by the timing resolution of the detection apparatus, results in a certain probability to accidentally detect two uncorrelated photons in coincidence, which do not belong to the same pair. Second, the statistical nature of the down-conversion process inherently generates multi-photon emissions within the coherence time of the photons. This also results in uncorrelated detection events at Alice and Bob. Furthermore, the finite coincidence time window leads to uncorrelated accidental coincidences from background light and intrinsic detector dark counts. In addition, imperfections and misalignment in the setup (source, polarization analysis, etc.) introduce systematic errors. In the following, the errors from uncorrelated detection events are characterized by the accidentals visibility $V_{acc}$ and the systematic errors by the system visibility $V_{sys}$. The total correlation visibility is given by $V_{tot}=V_{sys}\cdot V_{acc}$.
\par
The effect of these error sources on the secure key rate are analyzed analytically within a model devised by X.-F. Ma \emph{et al.} \cite{Ma07}, which assumes pulsed operation of the SPDC source. The input parameters for the model are the photon-pair generation rate at the source, the ratio between the coincidence and single rates at Alice and Bob including detector efficiencies and the system visibility $V_{sys}$. The model yields an error probability and a secure key gain per pump pulse as a function of total two-photon attenuation in a QKD experiment. A lower bound for the final secure bit rate per pulse $R$ is then calculated using Koashi and Preskill's security analysis \cite{Koashi03} via
\begin{equation}\label{koashipreskill}
    R\geq \frac{1}{2}\{P_c[1-f(q)H_2(q)-H_2(q)]\}.
\end{equation}
Here, $P_c$ is the coincidence detection probability between Alice and Bob per pump pulse, $\frac{1}{2}$ is the basis reconciliation factor and $H_2$ is the binary entropy function \begin{equation}\label{entropy}
    H_2(x)=-x\log_2(x)-(1-x)\log_2(1-x).
\end{equation}
The correction factor $f(q)$ accounts for the fact that practical error-reconciliation protocols in general do not perform ideally at the Shannon limit. Instead of assuming $f(q)\approx1$, we used realistic values for the applied bidirectional error correction protocol CASCADE \cite{Brassard94}. Furthermore, for our case of a cw-type SPDC source, the model was adapted to yield a probability per coincidence time window instead of a probability per pump pulse. We want to remark that other error-reconciliation protocols (e.g. WINNOW \cite{Buttler03}) promise to perform more efficiently, and it will be the subject of future investigation whether it might carry advantages also for our specific experimental environment.
\par
Note that Equation (\ref{koashipreskill}) gives the final secure key rate in the limit of infinite key lengths. However, in a practical implementation the secure key is obtained via error correction and privacy amplification on a finite key, which will further reduce the secure key rate to some extend (see \cite{Ma07,scarani08} for more details). For simplicity, we will restrict the analysis of our experiments to the infinite bounds.

\section{Description of the experiments and results}

\begin{figure}[!htb]
\includegraphics[width=130mm]{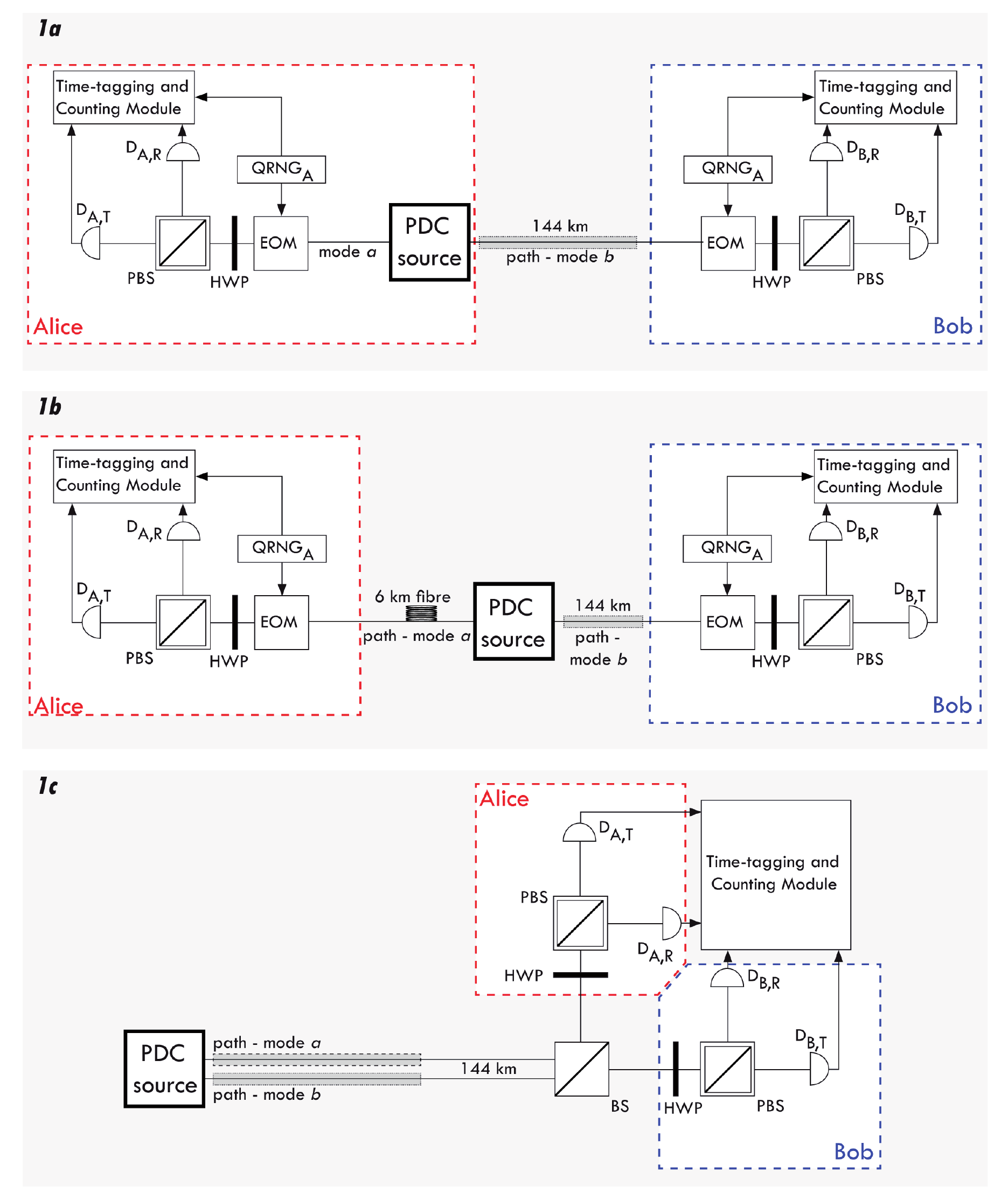}\\
\caption{An illustration of the three setups used for the entanglement-based quantum key distribution experiments. The entangled photon source (PDC source) generates entangled photon pairs that were sent via optical fibers and free-space optical links to Alice and Bob, respectively. There they were analyzed and detected, using four avalanche photo detectors D$_{A,T}$, D$_{A,R}$, D$_{B,T}$ and D$_{B,T}$. \textbf{\emph{a})} Alice's photon was analyzed directly at the source after 1 meter of optical fiber and Bob's photon was sent through a 144 km free-space channel. The total two-photon attenuation was 35~dB.  \textbf{\emph{b})} Alice's photon is now sent through a 6 km fiber, resulting in a total two-photon loss of 58~dB. In the scenarios \textbf{\emph{a})} and \textbf{\emph{b})} each analyzer module consisted of a polarizing beam splitter cube (PBS) and an electro optical modulator (EOM) to switch between the complementary bases. The EOMs were triggered by independent quantum random number generator. \textbf{\emph{c})} Both photons were sent through the 144 km free-space channel to one common receiver. There they were split up with a 50/50 beam splitter (BS) and guided to Alice and Bob, who could adjust their analyzing bases, using a half wave plate (HWP) and a PBS.}\label{Setupfigure}
\end{figure}

We implemented three different experimental QKD scenarios (see Figure \ref{Setupfigure}). In all three experiments, one photon of an entangled pair was sent to Bob via a 144 km free-space link, established between the islands of La Palma and Tenerife. The first and the second experiment were both asymmetrical with respect to the different channel losses for Alice and Bob. In the first experiment the SPDC source was placed at Alice (\emph{source at Alice}) and one photon of an entangled pair was measured directly at the source. In the second experiment (\emph{source asymmetric in between Alice and Bob}), Alice's photon was sent through a 6 km single-mode fiber before it was analyzed. In the third scenario both photons were sent via the 144 km free-space link to a common receiver where they were split up and analyzed separately. This can be seen as an effective realization of the \emph{source in the middle} scheme, since we have equal channel losses for Alice's and Bob's photons.
\par
To get an overview about the different scenarios and the corresponding results, please refer to Table \ref{summary}. A detailed discussion will be given in the next sections.

\begin{table}[!htb]
\begin{center}
\begin{tabular}{|c|c|c|c|c|c|}
\hline
\textbf{Scenario }               & \textbf{Attn.} & \textbf{local pair rate} & $\textbf{V}_{\textbf{tot}}$ & \textbf{QBER}  & \textbf{secure key rate}\\
 source...& [dB] &  [MHz] & [\%] & [\%] & [bits/s]\\
\hline\hline
\hline\hline
... at Alice   & 35 & 0.55  	& 86.2  	& 6.9 	& 24  \\
(Figure \ref{Setupfigure}a)   &  &  	&   	&  	&   \\
... asymmetric in between  & 58 & 2.5  	& 86.2  	& 6.8 	& 0.6 \\
(Figure \ref{Setupfigure}b)   &  &  	&   	&  	&   \\
... in the middle   & 71 & 1 		& 92 	& 4  		& 0.02  \\
(Figure \ref{Setupfigure}c)   &  &  	&   	&  	&   \\
\hline
\end{tabular}
\end{center}
\caption{A summary of the parameters for the three different experimental scenarios and the corresponding results, i.e., the total two-photon attenuation, the locally detected coincidence rate, the total visibility $V_{tot}$, the quantum bit error ratio QBER and the finally obtained secure key rate.} \label{summary}
\end{table}

\clearpage

\subsection{Source at Alice}\label{35dB}
The experimental situation is depicted in Figure \ref{Setupfigure}a. The SPDC source \cite{Fedrizzi07b} was located in La Palma and generated photon pairs in the entangled state (\ref{psiminus}). The photons in modes \emph{a} and \emph{b} were coupled into single-mode fibers, Alice's photon in mode \emph{a} was analyzed and detected locally after only 1 meter of single-mode fiber, while the photon in mode \emph{b} was sent through a 144 km free-space channel to Tenerife. There it was collected by the 1 meter diameter telescope of the optical ground station (OGS) operated by the European Space Agency ESA and analyzed by Bob.
\par
Each polarization analyzer consisted of an electro optical modulator (EOM), a polarizing beam splitter (PBS) and two single-photon avalanche diodes. Triggering the EOMs by independent quantum random number generators, the analyzer modules randomly switched between the complementary analyzing bases $|H,V\rangle$ and $|P,M\rangle$ as required for the BBM92 protocol \cite{Bennett92}. At Alice and Bob, every detection event (including arrival time, detector channel and EOM setting information) was recorded onto local computer hard disks, using time-tagging units disciplined by the global positioning system (GPS) time standard. Note that using active analyzers which are triggered by a quantum random number generator represents a security advantage over passive QKD systems, because it prevents an eavesdropper from applying certain side-channel attacks \cite{Ma07,Makarov05} (e.g. faked states attack) .
\par
In the first scenario implemented, the free-space channel attenuation for photons in mode \emph{b} was measured to be approximately 32~dB on average (including all optical elements), while only half of the locally measured photons (3~dB) in mode \emph{a} were lost in Alice's analyzer module. The total two-photon attenuation was therefore 35~dB.

The SPDC source generated entangled photon pairs at an estimated rate of 7~MHz, limited by the peak count rate of Alice's detector system. After single-mode fiber coupling, 550000 coincidences were observed locally, corresponding to a combined coupling and detection efficiency of 28\%. The darkcount rate at Alice was 500~Hz while Bob's detectors showed an average of 1200~Hz. For these parameters and a coincidence window of $\tau_c=1.5~\mbox{ns}$, theory predicts an upper bound of $V_{th}=94.1\%$ for the total visibility, which includes the initially measured system visibility $V_{sys}=96\%$ as well as the background and multi-pair emission limited visibility of $V_{acc}=98\%$ . However, the actual visibility of the transmitted entangled state was measured to be $V_{tot}=86.2\%$ on average. The discrepancy to $V_{th}$ was most probably caused by a polarization drift in the fiber connecting the source with the transmitter telescope during the measurements.
\par
\begin{figure}[!htb]
\begin{center}
  \includegraphics[width=120mm]{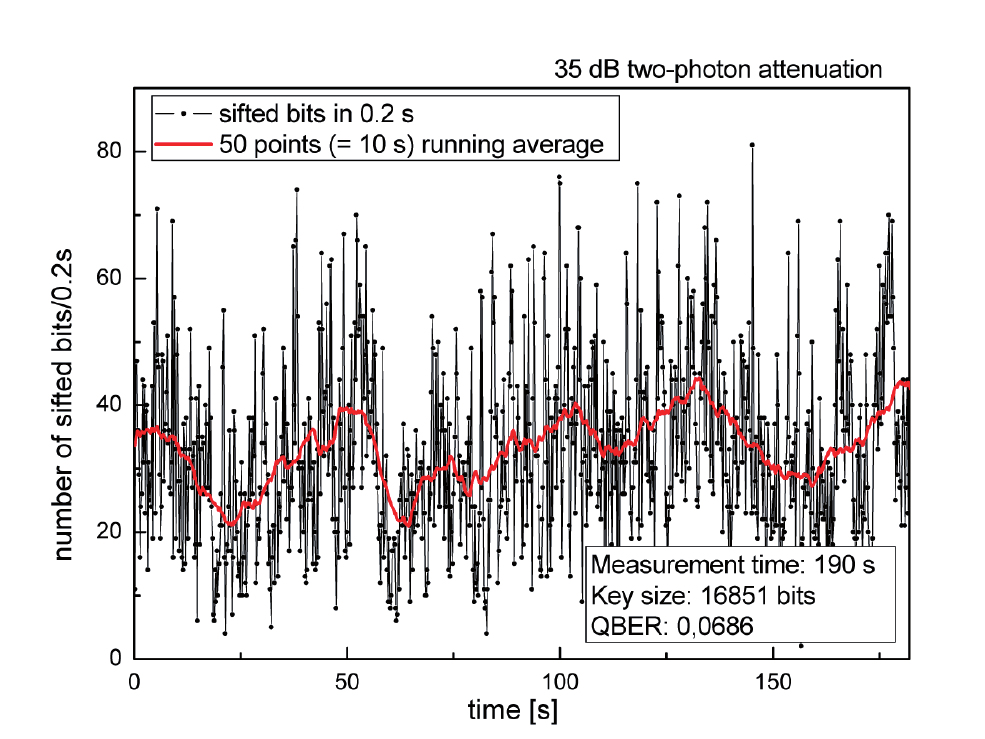}\\
  \caption{The result of a typical QKD measurements as shown in Figure \ref{Setupfigure}a, where Alice is located at the source and with a two-photon attenuation of 35~dB. The sifted key rates in 0.2 s are plotted versus measurement time. The strong intensity fluctuations through the free-space link are reflected in the fluctuations of the key rates. Accumulating three such measurements and applying Koashi and Preskill security analysis yielded an average secure key rate of approximately 24 bit/s.}\label{keyrates35dB}
\end{center}
\end{figure}

The result of a typical measurement is shown in Figure \ref{keyrates35dB}. In total, three measurements were performed and sifted keys containing 11024~bits (130 s
integration time), 13642~bits (190 s integration time) and 16851~bits (190 s integration time), respectively, were obtained. During a measurement run, the free-space link usually undergoes strong atmospheric turbulence, resulting in a time dependent two-photon attenuation. This is reflected by the time-resolved sifted key rates (see Figure \ref{keyrates35dB}). The QBERs for these measurement runs of 6.6\%, 7.3\% and 6.9\%, respectively, were obtained by comparing the sifted keys and are in good agreement with the measured overall visibility. Finally, applying Equation (\ref{koashipreskill}) with $f(q)\approx1.18$ yields an averaged secure key rate of approximately 24~bits/s. A comparison of all experimental data points to the theoretically calculated secure key rate as a function of overall link attenuation is shown in Figure \ref{theoryplusdatapoints}.

\subsection{Source asymmetric in between Alice and Bob}
The second scenario extends the \emph{source at Alice} scheme (see section \ref{35dB}), such that the photons in mode \emph{a} were delayed by 29.6~$\mu$s in a 6 km long single-mode fiber (see Figure \ref{Setupfigure}b). The attenuation of the fiber was measured to be 17~dB and during this particular measurement series, the free-space link attenuation was 38~dB. Combined with the 3~dB loss in Alice's analyzer, the overall two-photon attenuation was 58~dB.
For this experiment, we increased the output of the SPDC source to a pair generation rate of $32~\mbox{MHz}$ by operating at the maximum available pump laser power of 50~mW. Due to detector saturation, the fiber-coupled and locally detectable pair rate could only be extrapolated to be 2.5~MHz. In this situation, the initial system visibility was $V_{sys}=94\%$, a slight reduction compared to the 35~dB scenario, caused by the delay fiber. With the same coincidence window ($\tau_c=1.5~\mbox{ns}$) and darkcount rates as in the first experiment (500~Hz at Alice and 1200~Hz at Bob), the theoretic upper bound for this scheme turns out to be $V_{th}=88\%$ and the measured total visibility of the entangled state at the receiver was with $V_{tot}=86.2\%$ coincidentally the same as in the first scenario.
\par
\begin{figure}[!htb]
\begin{center}
  \includegraphics[width=110mm]{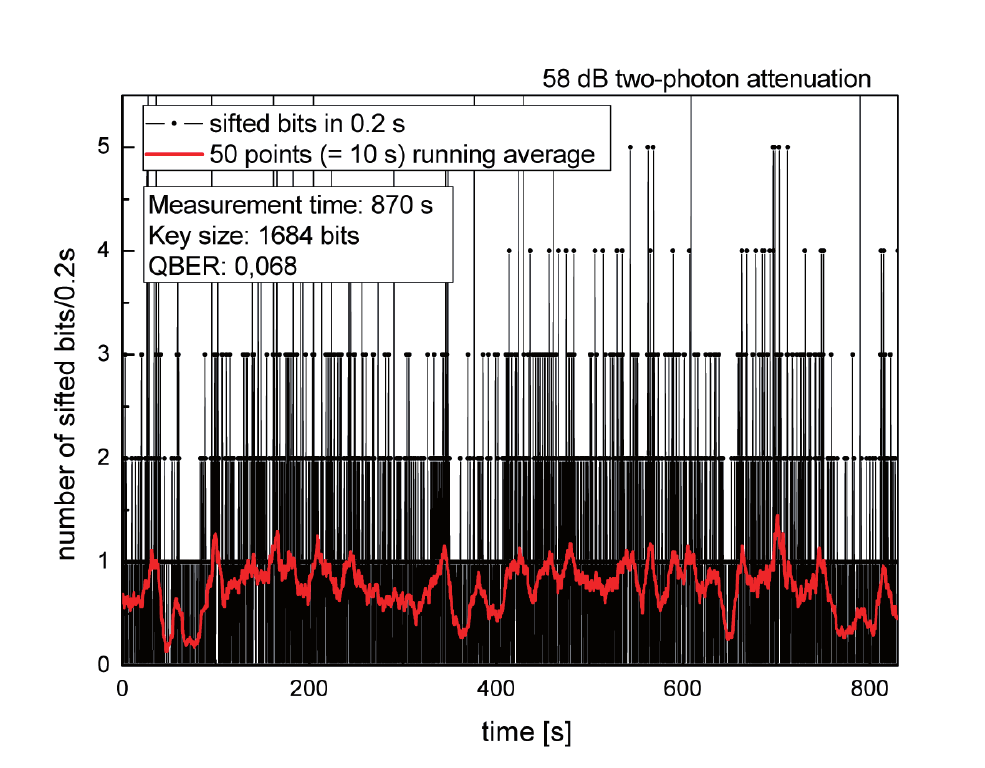}\\
  \caption{A typical measurement run for the scenario shown in Figure \ref{Setupfigure}b, where the source was arranged asymmetrically between Alice and Bob, resulting in a two-photon attenuation of 58~dB. Accumulating data for approximately 1400 s, an average secure key rate of 0.6~bits/s was obtained.}\label{keyrates58dB}
\end{center}
\end{figure}

A typical measurement result is depicted in Figure \ref{keyrates58dB}. In total, two sifted keys were obtained, containing 1107~bits (580 s integration time) and 1684~bits (880 s integration time). The corresponding QBERs were 6.9\% and 6.8\%, respectively, and a secure key rate of 0.6~bits/s could be obtained from Equation (\ref{koashipreskill}) with $f(q)\approx1.18$. For this scenario both the expected visibility and the key rate agree very well to the model (see Figure \ref{theoryplusdatapoints}).

\subsection{Source in the middle}
The experimental situation for the third, the \emph{source in the middle} scenario, is depicted in Figure \ref{Setupfigure}c. The entangled photons in mode \emph{a} and mode \emph{b} were coupled into single mode fibers, guided to two separate transmitter telescopes and sent through a 144 km free-space channel to one common receiver in Tenerife. The total two-photon attenuation was measured to be about 71~dB (including all optical components). For a detailed description of the setup please refer to \cite{fedrizzi09a}.

\begin{figure}[!htb]
\begin{center}
  \includegraphics[width=140mm]{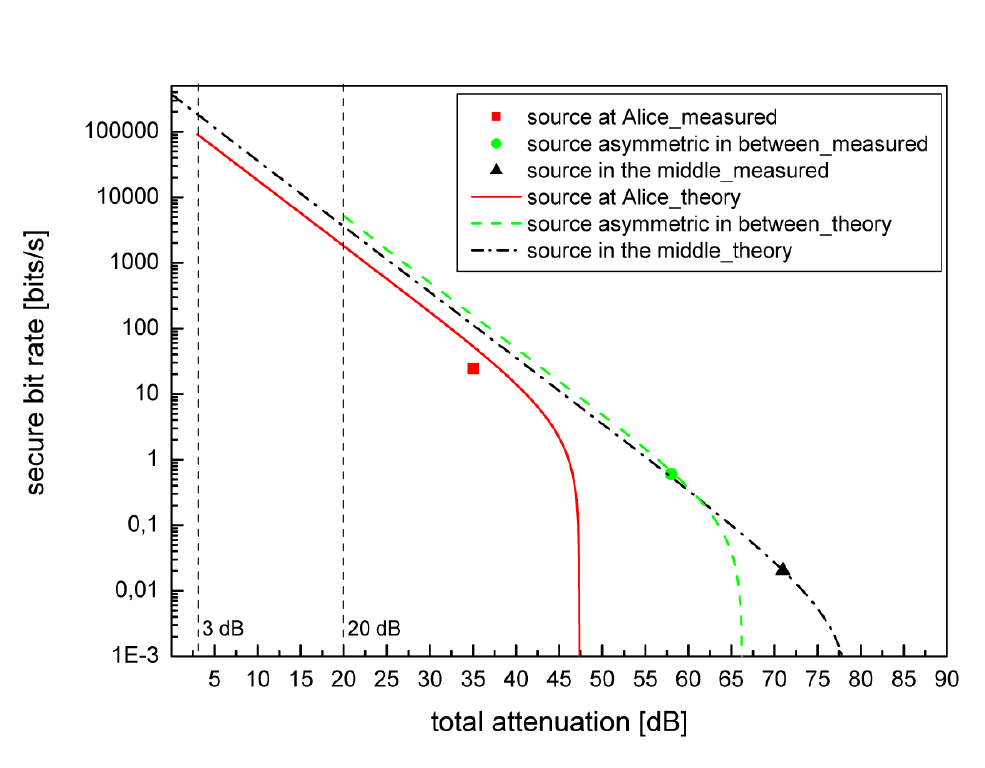}\\
  \caption{A comparison of the obtained results with the theoretical model described in the main text. The solid curve (\emph{source at Alice}) starts at a two-photon attenuation of 3~dB, which corresponds to the fixed loss in Alice's analyzer module. The dashed curve represents the scheme with the \emph{source asymmetrically between Alice and Bob}, where the attenuation for Alice's fiber channel together with the analyzer module was 20~dB. The dotted-dashed curve is predicted by our model for the \emph{source in the middle} scheme. The three experimentally obtained secure bit rates are depicted as the square, the circle and the triangle, respectively. It is easy to see that the data point for the \emph{source in the middle} scenario (triangle) can not be explained by the models for the asymmetric cases. Similarly, the data point of the experiment with the \emph{source asymmetrically between Alice and Bob} can not be explained by the model for the \emph{source at Alice} scheme. Thus the advantage of the symmetric scenario is clearly verified by our experimental results. Furthermore, these results also show that our system should be able to generate secure keys at high rates from 10~kbit/s at 15~dB and 100~kbit/s at 3~dB, a range typical for free-space links in metropolitan areas.
  }\label{theoryplusdatapoints}
\end{center}
\end{figure}

The source produced photon pairs at a rate of 10~MHz from which 3.3~MHz single photons and 1~MHz photon-pairs were detected locally. Both photons were sent via two telescopes over the 144 km free-space links to Tenerife. On average, 0.071 transmitted photon pairs/s could be detected, using a coincidence window of 1.25 ns. Each detector registered a background count rate of 400~Hz. Accumulating data for a total amount of 10800 seconds we measured an averaged visibility of the transmitted entangled state of $V_{tot}=92\%$ (with $V_{sys}=99\%$ and $V_{acc}=94\%$), which is very close to the theoretic upper bound of $V_{th}=91.7\%$ .
Based on these measurements we inferred that a QKD experiment employing a similar setup would have yielded a QBER of approximately 4\%. From the coincidence rate and the QBER-dependent performance of the classical key distillation protocols ($f(0.04)\approx1.16$), we estimate that our experiment would have yielded a final secure bit rate of approximately 0.02~bits/s (see Figure \ref{theoryplusdatapoints}). However, the implementation of a full QKD experiment was not possible, because only one receiver station and module was available in Tenerife.

\subsection{Clock synchronization}
As an additional feature, our coincidence search algorithm used
during the first two experiments could be utilized to
synchronize the individual time bases at Alice and Bob within 0.5 ns. For the third experiment such a synchronization was not necessary, because one and the same time-tagging system was used for the measurements.
\par
\begin{figure}[!htb]
\begin{center}
  \includegraphics[width=140mm]{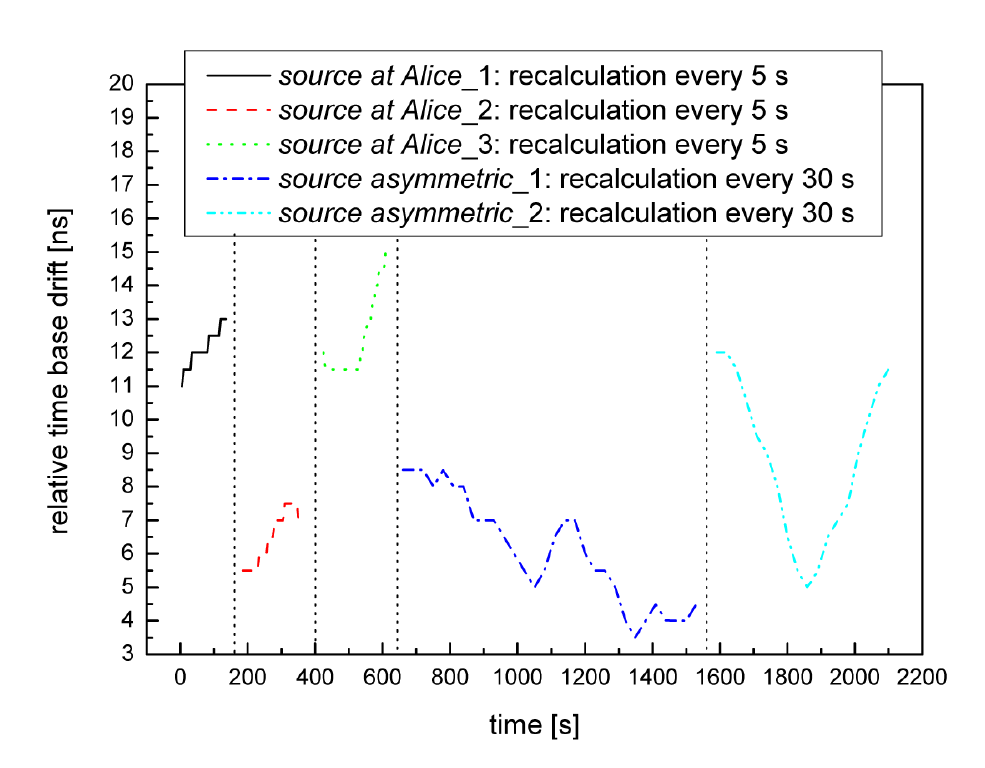}\\
  \caption{
This plot shows the relative drift between Alice's and Bob's local time-tagging systems that were directly disciplined by the
global positioning system. Due to the relative drift (vertical axis of the plot), the offset for the coincidence analysis was adapted by recalculating the cross-correlation. For the three measurements within the \emph{source at Alice} scheme the recalculation was performed in steps of 5 s (black, red and green curve), while for the two measurements within the \emph{source asymmetric in between Alice and Bob} scenario it was performed every 30 s (blue and light blue curve).
  }\label{gpsdrift}
\end{center}
\end{figure}

Coincidence events between Alice and Bob were identified by calculating the cross-correlation
function of the individual time-tagging data sets. A peak in the
cross-correlation function indicated the current time offset $\Delta
t$ between the time scales of the receiver units. Initially, Alice's and Bob's time bases were both disciplined by the GPS time standard. However, the two
individual GPS receivers exhibited a relative drift during a measurement run. By analyzing
the data in blocks of adjustable length, our software measured
and compensated for this relative drift with 0.5 ns resolution by
temporal alignment of the data blocks.
The data of the first experiment described were analyzed in blocks
of 5 s length, while the data obtained in the second experiment
were analyzed in blocks of 30 s length. The corresponding results concerning the relative drift of Alice's and Bob's time bases are depicted in Figure \ref{gpsdrift}.

\section{Conclusion}
We experimentally studied entanglement based QKD in three different high-attenuation scenarios on a 144~km free-space link between the two Canary Islands La Palma and Tenerife to verify the symmetry advantage of such an implementation. This involved placing the source directly at Alice, asymmetrically between Alice and Bob, and finally, symmetrically between the two parties. Our results demonstrate that in the symmetric case (\emph{source in the middle}) secure keys can be generated up to a channel attenuation of 70~dB, a regime in which asymmetric schemes fail. We conclude that entanglement-based QKD systems are the systems of choice for long distance quantum communication. Compared to the expected link attenuations in a low earth orbit (LEO) satellite to ground scenario of 30~dB, our results show, that entanglement-based systems are suitable for use in either a single-link (\emph{source at Alice}) or a two-link (\emph{source in the middle}) scenario. Our results imply that, using similar technology as in our experiment, it should readily be possible to implement QKD over distances of 300 km. This holds also true for fiber-based QKD, as there the attenuation is typically of the order of 0.2~dB/km, resulting in similar total attenuations.

\section{Acknowledgements}
The authors wish to thank F. Sanchez (Director IAC) and A. Alonso (IAC), T. Augusteijn, C. Perez and the staff of the Nordic Optical Telescope (NOT), J. Kuusela, Z. Sodnik and J. Perdigues of the Optical Ground Station (OGS) as well as J. Carlos and the staff of the Residence of the Observatorio del Roque de Los Muchachos for their support at the trial sites. This work was supported by ESA (contract number 18805/04/NL/HE), the Austrian Science Foundation (FWF) under project number SFB1520, the Doctoral Program CoQuS, the European project QAP, the Austrian Research Promotion Agency (FFG) through the Austrian Space Program ASAP, and the DTO-funded U.S. Army Research Office within the QCCM program. We also thank N. L\"utkenhaus for his valuable input.

\section{References}
\bibliographystyle{unsrt}   

\begin{thebibliography}{10}

\bibitem{Ma07}
X.~Ma, C.-H.~F. Fung, and H.-K. Lo.
\newblock Quantum key distribution with entangled photon sources.
\newblock {\em Phys. Rev. A}, 76:012307, 2007.

\bibitem{Hwang03}
W.-Y. Hwang.
\newblock Quantum key distribution with high loss: Toward global secure
  communication.
\newblock {\em Phys. Rev. Lett.}, 91(5):057901, Aug 2003.

\bibitem{peng07}
C.-Z. Peng, J.~Zhang, D.~Yang, W.-B. Gao, H.-X. Ma, H.~Yin, H.-P. Zeng,
  T.~Yang, X.-B. Wang, and J.-W. Pan.
\newblock Experimental long-distance decoy-state quantum key distribution based
  on polarization encoding.
\newblock {\em Physical Review Letters}, 98(1):010505, 2007.

\bibitem{hasegawa07}
J.~Hasegawa, M.~Hayashi, T.~Hiroshima, A.~Tanaka, and A.a Tomita.
\newblock Experimental decoy state quantum key distribution with unconditional
  security incorporating finite statistics.
\newblock {\em arXiv:0705.3081}, 2007.

\bibitem{Schmitt07}
T.~Schmitt-Manderbach, H.~Weier, M.~Fürst, R.~Ursin, F.~Tiefenbacher,
  T.~Scheidl, J.~Perdigues, Z.~Sodnik, C.~Kurtsiefer, J.~G. Rarity,
  A.~Zeilinger, and H.~Weinfurter.
\newblock Experimental demonstration of free-space decoy-state quantum key
  distribution over 144 km.
\newblock {\em Phys. Rev. Lett.}, 98:010504, 2007.

\bibitem{Ralph99}
T.~C. Ralph.
\newblock Continuous variable quantum cryptography.
\newblock {\em Phys. Rev. A}, 61(1):010303, Dec 1999.

\bibitem{Hillery00}
M.~Hillery.
\newblock Quantum cryptography with squeezed states.
\newblock {\em Phys. Rev. A}, 61(2):022309, Jan 2000.

\bibitem{ralph00}
T.~C. Ralph.
\newblock Security of continuous-variable quantum cryptography.
\newblock {\em Phys. Rev. A}, 62(6):062306, Nov 2000.

\bibitem{lodewyck07}
J.~Lodewyck, M.~Bloch, R.~Garc\'{\i}a-Patr\'{o}n, S.~Fossier, E.~Karpov,
  E.~Diamanti, T.~Debuisschert, N.~J. Cerf, R.~Tualle-Brouri, S.~W. McLaughlin,
  and P.~Grangier.
\newblock Quantum key distribution over 25 km with an all-fiber
  continuous-variable system.
\newblock {\em Physical Review A (Atomic, Molecular, and Optical Physics)},
  76(4):042305, 2007.

\bibitem{qi07}
B.~Qi, L.-L. Huang, L.~Qian, and H.-K. Lo.
\newblock Experimental study on the gaussian-modulated coherent-state quantum
  key distribution over standard telecommunication fibers.
\newblock {\em Physical Review A (Atomic, Molecular, and Optical Physics)},
  76(5):052323, 2007.

\bibitem{Ekert91}
A.~K. Ekert.
\newblock Quantum cryptography based on bell's theorem.
\newblock {\em Phys.\ Rev.\ Lett.}, 67:661--663, 1991.

\bibitem{Bennett92}
C.~H. Bennett, G.~Brassard, and N.~D. Mermin.
\newblock {Q}uantum {C}ryptography {W}ithout {B}ell's {T}heorem.
\newblock {\em Phys.\ Rev.\ Lett.}, 68:557--559, 1992.

\bibitem{honjo08}
T.~Honjo, S.~W. Nam, H.~Takesue, Q.~Zhang, H.~Kamada, Y.~Nishida, O.~Tadanaga,
  M.~Asobe, B.~Baek, R.~Hadfield, S.~Miki, M.~Fujiwara, M.~Sasaki, Z.~Wang,
  K.~Inoue, and Y.~Yamamoto.
\newblock Long-distance entanglement-based quantum key distribution over
  optical fiber.
\newblock {\em Opt. Express}, 16:19118--19126, 2008.

\bibitem{magicQ}
MagicQ, Somerville, Massachusetts, USA.
\newblock http://www.magiqtech.com/.

\bibitem{quantique}
'Cerberis' and 'Clavis2' by ID Quantique, Geneva, Switzerland.
\newblock http://www.idquantique.com.

\bibitem{treiber09}
Alexander Treiber, Andreas Poppe, Michael Hentschel, Daniele Ferrini, Thomas
  Lorunser, Edwin Querasser, Thomas Matyus, Hannes Hubel, and Anton Zeilinger.
\newblock Fully automated entanglement-based quantum cryptography system for
  telecom fiber networks.
\newblock {\em arXiv:0901.2725v1}, 2009.

\bibitem{ursin07}
R.~Ursin, F.~Tiefenbacher, T.~Schmitt-Manderbach, H.~Weier, T.~Scheidl,
  M.~Lindenthal, B.~Blauensteiner, T.~Jennewein, J.~Perdigues, P.~Trojek,
  B.~Oemer, M.~Fuerst, M.~Meyenburg, J.~Rarity, Z.~Sodnik, C.~Barbieri,
  H.~Weinfurter, and A.~Zeilinger.
\newblock Entanglement-based quantrum communication over 144 km.
\newblock {\em Nature Physics}, 3:481 -- 486, 2007.

\bibitem{Bennett92d}
C.~H. Bennett, F.~Bessette, G.~Brassard, L.~Savail, and J.~Smolin.
\newblock {\em J. Cryptology}, 5:3, 1992.

\bibitem{jacobs96}
B.~C. Jacobs and J.~D. Franson.
\newblock Quantum cryptography in free space.
\newblock {\em Optics Letters}, 21:1854--1856, 1996.

\bibitem{Buttler98a}
W.~T. Buttler, R.~J. Hughes, P.~G. Kwiat, G.~G. Luther, G.~L. Morgan, J.~E.
  Nordholt, C.~G. Peterson, and C.~M. Simmons.
\newblock Free-space quantum-key distribution.
\newblock {\em Phys. Rev.~A}, 57:2379--2382, 1998.

\bibitem{Buttler00a}
W.~T. Buttler, R.~J. Hughes, S.~K. Lamoreaux, G.~L. Morgan, J.~E. Nordholt, and
  C.~G. Peterson.
\newblock Daylight quantum key distrbution over 1.6 km.
\newblock {\em Phys. Rev. Lett.}, 84:5652--5655, 2000.

\bibitem{kurtsiefer02}
C.~Kurtsiefer, P.~Zarda, M.~Halder, H.~Weinfurter, P.~M. Gorman, P.~R. Tapster,
  and J.~G. Rarity.
\newblock A step towards global key distribution.
\newblock {\em Nature}, 419:450, 2002.

\bibitem{kurtsiefer02b}
C.~Kurtsiefer, P.~Zarda, M.~Halder, Ph.~M. Gorman, P.~R. Tapster, J.~G. Rarity,
  and H.~Weinfurter.
\newblock Long-distance free-space quantum cryptography.
\newblock {\em Proc. SPIE}, 4917:25--31, 2002.

\bibitem{Erven08}
C.~Erven, C.~Couteau, R.~Laflamme, and G.~Weihs.
\newblock Entangled quantum key distribution over two free-space optical links.
\newblock {\em Opt. Express}, 16(21):16840--16853, 2008.

\bibitem{peloso08}
M.~P. Peloso, I.~Gerhardt, C.~Ho, A.~Lamas-Linares, and C.~Kurtsiefer.
\newblock Daylight operation of a free space, entanglement-based quantum key
  distribution system.
\newblock {\em arXive}, 0812.1880v1, 2008.

\bibitem{Buttler98c}
W.~T. Buttler, R.~J. Hughes, P.~G. Kwiat, S.~K. Lamoreaux, G.G. Luther, G.~L.
  Morgan, J.~E. Nordholt, C.~G. Peterson, and C.M. Simmons.
\newblock Practical free-space quantum key distribution over 1 km.
\newblock {\em Phys.\ Rev.\ Lett.}, 81:3283--3286, 1998.

\bibitem{ursin08b}
R.~Ursin, T.~Jennewein, J.~Kofler, J.~M. Perdigues, and L.~Cacciapuoti \emph{et
  al.}
\newblock Space-quest: Experiments with quantum entanglement in space.
\newblock {\em International Aeronautical Congress Proceedings A2.1.3,
  arXiv:0806.0945}, 2008.

\bibitem{Bennett84}
C.~H. Bennett and G.~Brassard.
\newblock Quantum cryptography.
\newblock In {\em Proceedings of {IEEE} International Conference on Computers,
  Systems, and Signal Processing, Bangalore, India}, page 175, New York, 1984.
  IEEE.

\bibitem{Koashi03}
Masato Koashi and John Preskill.
\newblock Secure quantum key distribution with an uncharacterized source.
\newblock {\em Phys. Rev. Lett.}, 90(5):057902, Feb 2003.

\bibitem{Brassard94}
G.~Brassard and L.~Salvail.
\newblock Eurocrypt '93: Workshop on the theory and application of
  cryptographic techniques on advances in cryptology.
\newblock In {\em Lecture Notes in Computer Science}, volume 765, pages
  410--423, New York, 1994. Springer.

\bibitem{Buttler03}
W.~T. Buttler, S.~K. Lamoreaux, J.~R. Torgerson, G.~H. Nickel, C.~H. Donahue,
  and C.~G. Peterson.
\newblock Fast, efficient error reconciliation for quantum cryptography.
\newblock {\em Phys. Rev. A}, 67(5):052303, May 2003.

\bibitem{scarani08}
V.~Scarani, H.~Bechmann-Pasquinucci, N.~J. Cerf, M.~Dusek, N.~Luetkenhaus, and
  M.~Peev.
\newblock The security of practical quantum key distribution.
\newblock {\em arXiv:0802.4155v2 [quant-ph]}, 2008.

\bibitem{Fedrizzi07b}
A.~Fedrizzi, T.~Herbst, A.~Poppe, T.~Jennewein, and A.~Zeilinger.
\newblock A wavelength-tunable, fiber-coupled source of narrowband entangled
  photons.
\newblock {\em Optics Express}, 15:15377--15386, 2007.

\bibitem{Makarov05}
V.~Makarov and D.~R. Hjelme.
\newblock Faked states attack on quantum cryptosystems.
\newblock {\em Journal of Modern Optics}, 52:691--705, 2005.

\bibitem{fedrizzi09a}
A.~Fedrizzi, R.~Ursin, T.~Herbst, M.~Nespoli, R.~Prevedel, T.~Scheidl,
  F.~Tiefenbacher, T.~Jennewein, and A.~Zeilinger.
\newblock High-fidelity transmission of entanglement over a high-loss freespace
  channel.
\newblock \emph{Nature Physics}, doi: 10.1038/nphys1255, 2009.

\end{thebibliography}

\end{document}